\documentclass[aps,prd,showpacs,twocolumn,groupedaddress]{revtex4}
\usepackage{dcolumn}
\usepackage{graphicx}
\usepackage{amsmath,amssymb}

\providecommand{\h}{h}

\begin{document}

\title{Exotic Meson Decay Widths using Lattice QCD}

\author{M. S. Cook}
\email[]{mcook003@fiu.edu}
\author{H. R. Fiebig}
\email[]{fiebig@fiu.edu}

\thanks{This material is based on work supported by the U.S. National Science 
        Foundation under Grant No. PHY-0300065 and upon resources provided by the 
        Lattice Hadron Physics Collaboration through the SciDac program of the 
        U. S. Department of Energy.}

\affiliation{Department of Physics, Florida International University, \\ 
        Miami, Florida, USA 33199}

\date{\today}

\begin{abstract}
A decay width calculation for a hybrid exotic meson $h$, with $J^{PC}=1^{-+}$, 
is presented for the channel $h\rightarrow \pi a_1$. This quenched lattice QCD simulation
employs L{\"u}scher's finite box method. Operators coupling to the $h$ and $\pi a_1$ states are 
used at various levels of smearing and fuzzing, and at four quark masses. Eigenvalues of the corresponding correlation matrices yield energy spectra that determine scattering phase shifts for a discrete set of relative $\pi a_1$ 
momenta. Although the phase shift data is sparse, fits to a Breit-Wigner model
are attempted, resulting in a decay width of about 60 MeV when averaged over two lattice sizes.
\end{abstract}


\maketitle

\section{Introduction}
Hybrid mesons are quark-antiquark pairs having valence gluons as a structural component.
In some cases their quantum numbers are not accessible with quark models, and they
are therefore called exotic. Examples of these exotics are the $J^{PC}=0^{+-},1^{-+},\:\mbox{and}\;\,2^{+-}$ mesons.
Because these mesons contain valence gluons 
their verification is one of the signature tests of Quantum Chromodynamics (QCD).

Efforts to determine properties of these hybrid exotic states 
are unsettled from both experimental and theoretical viewpoints~\cite{Barnes:2003vy,Barnes:2000vn,Barnes:2000ns}.
The experimental efforts date back over a
decade~\cite{Alde:1988bv,Thompson:1997bs,Abele:1998gn,Adams:1998ff,Chung:1999we,Ivanov:2001rv,Dzierba2005},
and currently, considerable resources are being devoted to their future study.
The Jefferson Lab GlueX experimental program is designed to investigate exotic states. 
Hybrid meson studies are also part of the COMPASS experiment at CERN 
and the CLEOc program at Cornell. 

Attempts to calculate decay widths of hybrid mesons have been made using the bag model \cite{Barnes:1982tx},
the quark model extended by gluon flux tube degrees of freedom \cite{Barnes:1995hc,Close:1995hc}
and in lattice QCD \cite{McNeile:2006bz,McNeile:2002az}.
 
Calculating properties of resonances using Euclidean lattice QCD simulations is
not straightforward for a variety of reasons.
On a finite lattice all states are bound, and furthermore, the lattice total energy of a 
two-hadron state in a decay channel is typically larger than the energy of the
original hadron thus preventing decay.
Aspects of these points have been discussed by Michael~\cite{Michael:1989mf,Michael:2005},
DeGrand~\cite{Degrand:1991dg}, L{\"u}scher~\cite{Luscher:1991cf}, and by Lellouch
and L{\"u}scher~\cite{Lellouch:2000pv}.

Thus, lattice work on hybrid mesons has concentrated mainly on their mass spectrum. It can be
roughly classified in terms of heavy quark systems using static quarks with (excited) glue treated
in the Born-Oppenheimer approximation~\cite{Juge:1999ie,Juge:1998nc,Perantonis:1990dy}, using
non-relativistic QCD~\cite{Burch:2001tr,Drummond:1999db,Manke:1998tb}, and studies using actions
with both quenched and unquenched quark
dynamics~\cite{McNeile:1998cp,Bernard:1997ib,Bernard:1997bg,Lacock:1998be,Lacock:1998an,Lacock:1997ny,Lacock:1996vy,Bernard:2003cw}.
Lattice work on hadron resonances, of which there is very little at this time,
has recently been reviewed by Michael \cite{Michael:2005}. 

Hybrid meson decay widths have been studied on the lattice in the heavy 
quark limit~\cite{McNeile:2002az} and, recently, for light quarks \cite{McNeile:2006bz}.
In those studies the time dependence (slope) of a normalized transition matrix element
computed on the lattice is related to a decay width via Fermi's golden rule \cite{McNeile:2002fh}.
For this approach to work the lattice parameters have to be such that the resonance mass
comes out close to the threshhold of the decay channel.

We here choose to extract decay widths using L{\"u}scher's finite box
method~\cite{Luscher:1991cf,Luscher:1991ux,Cook:2006mc}. In principle L{\"u}scher's method is
rigorous:
The two-particle energy spectrum in a finite periodic box is related to continuum
elastic scattering amplitudes. The spectra allow calculation
of the scattering phase shifts at a discrete set of momenta owing to exact formulae derived by 
L{\"u}scher. A decay width can then be extracted by fitting a Breit-Wigner function,
provided a resonant state is actually present.
The applicability of this method to extract scattering phase shifts has been demonstrated
for the O(3) non-linear sigma model in 1+1 
dimensions~\cite{Luscher:1990ck}, the O(4) non-linear sigma model in 3+1 
dimensions~\cite{Gockeler:1994rx,Zimmermann:1993kx}, meson-meson
scattering in 2+1 dimensions using QED~\cite{Fiebig:1994qi}, and resonance scattering of two coupled 
Ising systems~\cite{Gat93b,Gat93a}.

Application of L{\"u}scher's method to our desired goal requires a set of operators that couple to
the hybrid meson state and to appropriate two-meson systems matching a decay channel. The exotic
$1^{-+}$ meson can decay into $\pi b_1$, $\pi f_1$, and $\pi a_1$, which are relative S-wave
channels. Other decays are possible, but those involve relative P-waves, where the relative momentum is
at least $2\pi/L$, and thus give rise to large upward energy shifts which makes the
simulation more difficult \cite{Degrand:1991dg,McNeile:2002fh,Loft:1988sy}.
We note that experimentally \cite{PDBook:2004} the three mesons $b_1(1235)$, $a_1(1260)$, $f_1(1285)$
are close in mass, and that the $a_1$ is a vector meson with $C=+$ so that it combines
naturally with $C=0$ of the (neutral) pion to the required charge conjugation, $C=+$, of the
exotic meson.    
Therefore, in this study, we model our two-hadron operator after the $\pi a_1$ decay channel. 

\section{\label{sec:lat}Lattice Parameters}
Obtaining excited state spectra from correlation matrices that involve
two-hadron operators presents a numerical challenge in itself. 
On top of this more analysis steps are required to obtain a decay width,
which essentially amounts to extracting a ``derivative'' quantity from the simulation.
While statistical errors, and also discretization errors to some extent, can be easily
controlled this is not true for the systematic uncertainties indigenous to a task of
this kind. For this reason we have refrained from using large computing
resources and therefore employed a very simple lattice action and moderate lattice
sizes and pion masses.

Simulations were performed using the Wilson gauge field action and Wilson 
fermions in quenched approximation on anisotropic lattices.
We will present results
from 200 gauge configurations on $12^3\times24$ and $10^3\times24$ lattices. 

The definition of the coupling parameters used for the Wilson gauge action 
\begin{equation}\label{puregluon}
S_{g}[U]= \sum_{x}\sum_{1\le \mu < \nu \le 4}\beta_{\mu\nu}\biggl(1-\frac{1}{3}\mbox{\,Re\,Tr\,}(U_{\mu\nu}(x))\biggr)
\end{equation}
is given by
\begin{equation}
\beta_{\mu\nu}=\beta \frac{a_1 a_2 a_3 a_4}{(a_\mu a_\nu)^2}\,,
\end{equation}
where $\mu\nu$ denotes plaquette planes and $a_1=a_2=a_3=:a_s$ is the spatial
and $a_4=:a_t$ the temporal lattice constant. We have chosen the (bare) anisotropy
$\xi=a_s/a_t=2$ and $\beta=6.15$ for the global coupling parameter.
Parameters for the anisotropic fermion matrix
\begin{eqnarray}\label{quarkm}
Q(x,y)=\mathbf{1}\,\delta_{x,y}-\sum_{\mu}\kappa_{\mu}\biggl((1-\gamma_{\mu})U_{\mu}(x)\delta_{x+\hat{\mu},y} &&\nonumber \\
+(1+\gamma_\mu)U^{\dagger}_{\mu}(y)\delta_{x,y+\hat{\mu}}\biggr) &&
\end{eqnarray}
are given by
\begin{equation}\label{kappamu}
\kappa_{\mu}=\kappa\frac{4}{a_{\mu}\sum_{\lambda}\frac {1}{a_\lambda}}
\end{equation}
where $\kappa$ is a global hopping parameter.
With these conventions the relation of the latter to the (bare) Wilson quark mass
parameter $m_q$ is
\begin{equation}
\kappa=\frac{\sum_{\lambda}\frac{1}{a_{\lambda}}}{8(m_q+\sum_{\lambda}\frac{1}{a_{\lambda}})}\,,
\end{equation}
which identifies $\kappa_c=0.125$ as the critical value.

Correlation functions for mesons from standard local operators $\pi\sim\gamma_5$,
$\rho\sim\gamma_i$, and $a_1\sim\gamma_i\gamma_5$, $i=1,2,3$ were
constructed as a matter of course employing three iterations of quark field
smearing \cite{Alexandrou:1994ti} and gauge field fuzzing \cite{Alb87a}.
Also, within this setting, we adopted the hybrid meson operator proposed
in \cite{Bernard:1997ex} with magnetic type gluons
\begin{equation}\label{hyopert} 
O_{\h}(t)= \sum_{1\le i < j \le 3}
\sum_{\vec{x}}\bar{d}_{a}(\vec{x}t) \gamma_{i} u_{b}(\vec{x}t) \,[F^{ab}_{ij}(\vec{x}t)- F^{\dagger ab}_{ij}(\vec{x}t)]\,,
\end{equation}
where $a,b$ denote color indices and $F_{ij}(x)$ is a product of SU(3) link matrices arranged in a
clover pattern
\begin{eqnarray}\label{gluonfields}         
\lefteqn{F_{\mu\nu}(x) = U_{\mu}(x)U_{\nu}(x+\hat{\mu})U^{\dagger}_{\mu}(x+\hat{\nu})U^{\dagger}_{\nu}(x)\notag}\\
&+& U_{\nu}(x)U^{\dagger}_{\mu}(x-\hat{\mu}+\hat{\nu})U^{\dagger}_{\nu}(x-\hat{\mu})U_{\mu}(x-\hat{\mu})\notag\\
&+& U^{\dagger}_{\mu}(x-\hat{\mu})U^{\dagger}_{\nu}(x-\hat{\mu}-\hat{\nu})U_{\mu}(x-\hat{\mu}-\hat{\nu})
U_{\nu}(x-\hat{\nu})\notag\\
&+& U^{\dagger}_{\nu}(x-\hat{\nu})U_{\mu}(x-\hat{\nu})U_{\nu}(x+\hat{\mu}-\hat{\nu})U^{\dagger}_{\mu}(x)\,,
\end{eqnarray}
which is used in the spatial planes only. Under parity we have
${\cal P}O_{h}(t){\cal P}^{-1} = -O_{h}(t)$,
while for the charge neutral ($\bar{d}u\rightarrow \bar{u}u,\bar{d}d$ ) version $O_{h^{0}}(t)$ of
(\ref{hyopert}) under charge conjugation the derivation of 
${\cal C}O_{h^{0}}(t){\cal C}^{-1} = -O_{h^{0}}(t)$
relies on the presence of $F-F^\dagger$, specifically ${\cal C}F_{ij}(x){\cal C}^{-1}=F^\ast_{ij}(x)$.
We have enforced this relation in the simulation:
Observing that $S_g[U]=S_g[U^\ast]$ the configurations
$[U]$ and $[U^\ast]$ are equally probable. Thus with each $[U]$ in the ensemble of 200
configurations we also include $[U^\ast]$ and compute fermion propagators for both of those.
This strategy doubles the number of fermion propagators that need to be computed, however,
charge conjugation is now numerically exact, and this also appears to be the reason for an observed
noise reduction of simulation signals.

Meson masses were obtained at four values of the hopping parameter $\kappa$,
see Table~\ref{tab:hopping}.
A multiple mass inverter \cite{Frommer:1995ik} was used to compute propagators.
The resulting ground state masses, coming from three smearing iterations, sources set at $t=3$,
and effective mass function fits in the range $t=6\ldots 11$, are also
listed in Tab.~\ref{tab:hopping}.
\begin{table}[h]
\caption{\label{tab:hopping}List of hopping parameters $\kappa$ and the resulting
$\pi$, $\rho$, $a_1$ and $h$ meson masses in units of the temporal lattice constant $a_t$
for lattices $12^{3}\times24$ (upper table) and $10^{3}\times24$ (lower table).}
\begin{ruledtabular}
\begin{tabular}{ccccc}
${\kappa}$ & ${a_{t}m_{\pi}}$ & ${a_{t}m_{\rho}}$ & $ {a_{t}m_{a_1}}$ & ${a_{t}m_{\h}}$ \\
\colrule
0.140 & 0.53(4) & 0.55(3) & 0.65(4) & 0.63(25) \\
0.136 & 0.64(3) & 0.65(3) & 0.75(3) & 0.75(21) \\ 
0.132 & 0.75(3) & 0.75(3) & 0.86(3) & 0.85(26) \\
0.128 & 0.85(4) & 0.85(3) & 0.96(3) & 0.95(24) \\
\colrule
0.140 & 0.54(5) & 0.54(3) & 0.64(4) & 0.62(24) \\
0.136 & 0.65(3) & 0.62(3) & 0.75(3) & 0.71(23) \\
0.132 & 0.74(3) & 0.73(3) & 0.85(4) & 0.81(25) \\
0.128 & 0.85(4) & 0.83(3) & 0.96(3) & 0.91(22) \\
\end{tabular}
\end{ruledtabular}
\end{table}

In order to allow extrapolations to the small pion mass region it is useful 
to study the dependence of the computed $\rho$, $a_{1}$ and $\h$ masses on $M_\pi=a_{t}m_\pi$.
Predictions for this dependence may in principle come from chiral perturbation
theory \cite{BernsteinHolstein:1995}, and will depend on the baryon being studied.
For a baryon of mass $M=a_{t}m$ the expression
\begin{equation}\label{chiPT}
M \approx p_0+p_2\,M_{\pi}^2+p_3\,M_{\pi}^3+p_4\,M_{\pi}^4+q\,M_{\pi}^4\ln(M_\pi)
\end{equation}
contains a collection of terms typical for $\chi$PT inspired
models \cite{Smit:2002sm,Wright:2002wr}.
In the case of the hybrid exotic meson, for example, the authors of \cite{Hedditch:2005zf}
retain only the even polynomial in (\ref{chiPT}).
For the $\pi a_1$ decay channel, which is mostly relevant in this work,
no predictions for the dependence of the spectral masses, say $W$, on $M_\pi$ are
available. A three parameter model that reflects features of (\ref{chiPT}) is
\begin{equation}\label{fitln}
W=p+qx+r\ln (1+x) \quad\,\, \mbox{with} \,\,\, x=(a_{t}m_\pi)^2\,.
\end{equation}
The logarithmic term is purely heuristic. Its role is to provide curvature to the model,
just like the last three terms in (\ref{chiPT}) do while vanishing as $M_{\pi}\rightarrow 0$.
As it turns out this model yields fits that are on average optimal on our spectral
data for the combined $\h$ and $\pi a_1$ systems. 
Replacing the logarithmic term in (\ref{fitln}) with $x^{3/2}$ yields nearly identical
results.

\begin{figure}[h]
\includegraphics[height=82mm,angle=90]{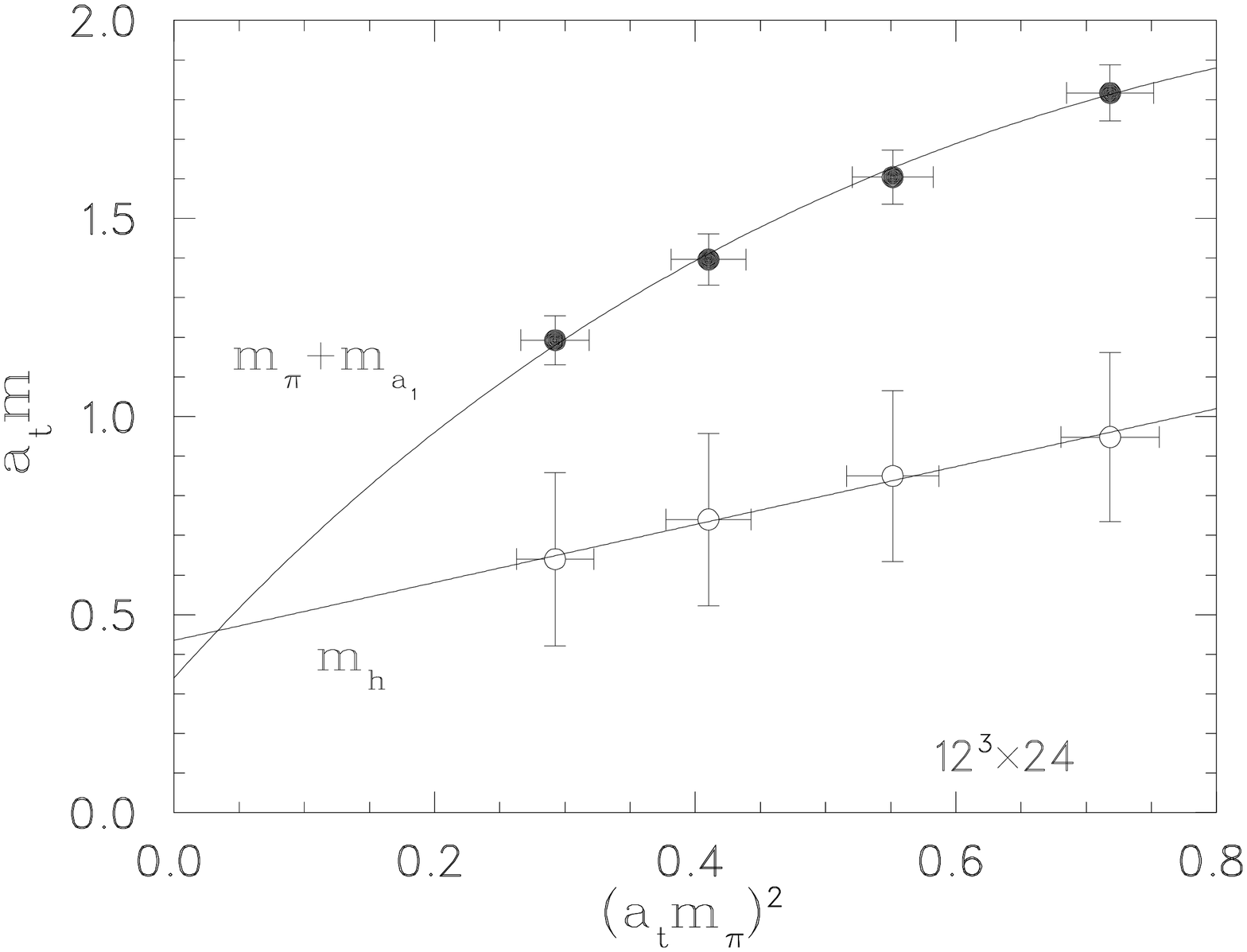}
\includegraphics[height=82mm,angle=90]{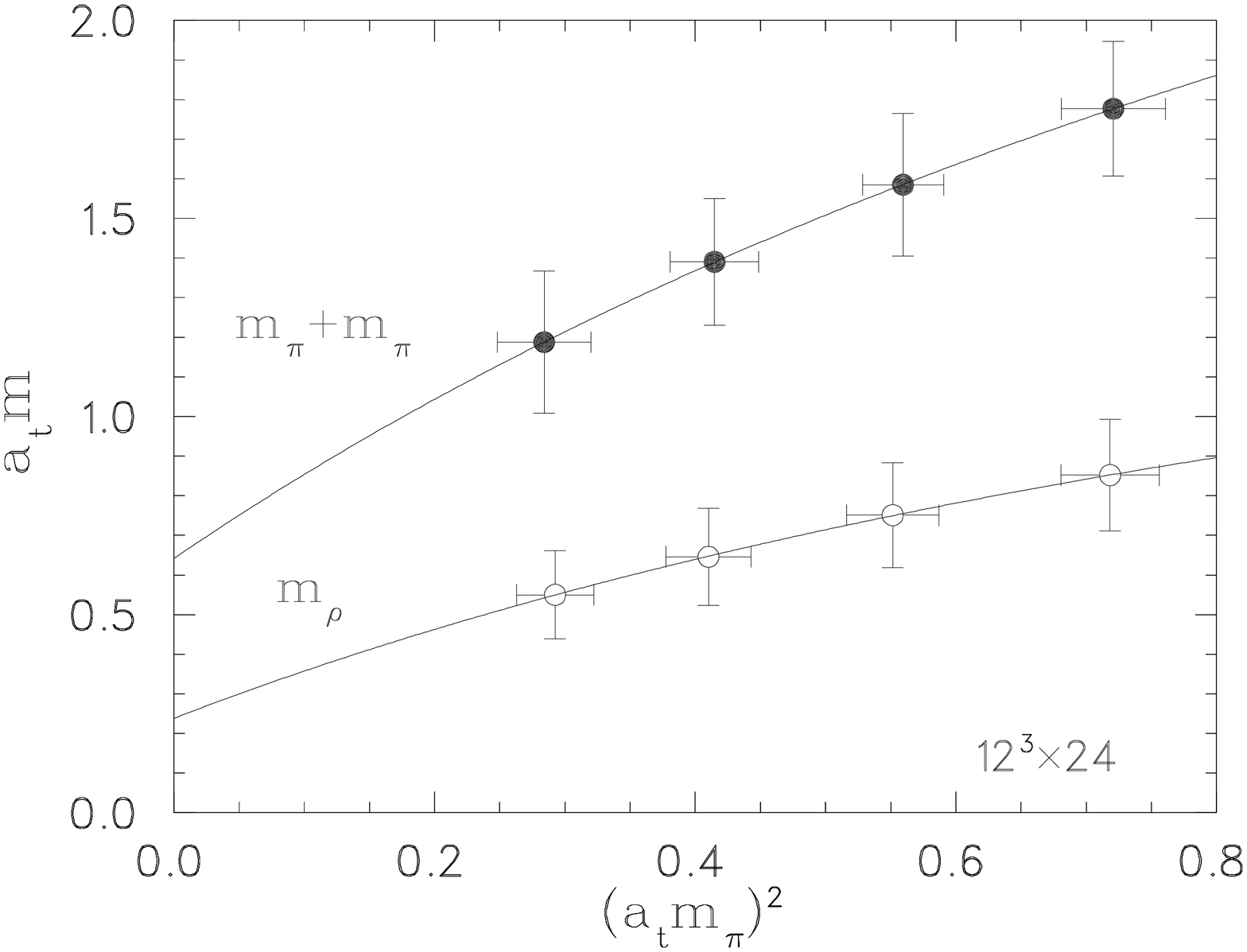}
\caption{\label{fig:effa1pionhyb}Combinations of masses, as indicated, obtained from single
meson operators versus the squared pion mass $x=(a_tm_\pi)^2$ and fits with the model
(\protect\ref{fitln}).
The relative S-wave $\pi+a_1$ mass (upper panel) reveals a level crossing near
zero pion mass through extrapolation.
The $\pi+\pi$ mass is shown with pions having lattice
momenta $\pm 2\pi/(a_sL)$.
The extrapolation of the $\rho$ meson mass to $x=0$ is used to set
the physical scale.}
\end{figure}

We consistently use (\ref{fitln}) to fit all masses emerging from the simulation.
Examples for the mesons listed in Tab.~\ref{tab:hopping} are shown in
Fig.~\ref{fig:effa1pionhyb}.

In particular the upper panel of Fig.~\ref{fig:effa1pionhyb} exhibits 
a level crossing between the hybrid meson mass and the $\pi+a_1$ mass,
assuming a relative S-wave for the latter. The level crossing emerges near
$x\simeq 0$ which is only reached through extrapolation. 
The lower panel illustrates the effect of P-wave~vs.~S-wave decay on the lattice.
The mass of the $\pi+\pi$ system is shown with pions having lattice
momenta $\pm 2\pi/(a_sL)$. Clearly a level crossing with the $\rho$ meson mass is
harder to achieve.

The extrapolated $\rho$ meson mass, at $x=0$, shall be used to
set the physical mass or length scale for this simulation. We obtain
$a_t=0.33(5)\,\mbox{GeV}^{-1}\,=0.07(1)\,\mbox{fm}$ ($m_{\rho}=776\,\mbox{MeV}$).
If the $a_1$ meson is used instead the scale is
$a_t=0.30(3)\,\mbox{GeV}^{-1}\,=0.06(1)\,\mbox{fm}$ ($m_{a_1}=1230\,\mbox{MeV}$).
The above numbers are based on the $12^{3}\times 24$ lattice.
Unless otherwise indicated
we will quote results using the $\rho$ meson to set the scale.

As a sideline it is interesting to note that the level crossing seen in
Fig.~\ref{fig:effa1pionhyb}, using both of the above scales, thus occurs
within 1.35--1.49GeV, which overlaps with the experimental mass of the $\pi(1400)$
resonance, according to \cite{PDBook:2004}. Indeed the $\pi(1400)$ has the quantum
numbers $1^{-+}$ of the hybrid exotic meson.
This observation coincides with the findings of \cite{Hedditch:2005zf}.

\section{Correlation matrix}
The description of $h\rightarrow\pi+a_{1}$
requires an operator for the two-meson decay channel with suitable
quantum numbers. Consider
\begin{eqnarray}\label{piaone}
\lefteqn{O_{\pi^{+}a^{0}_{1};k,\vec{r}\,}(t)=\sum_{\vec{x}}\sum_{\vec{y}}\delta_{\vec{x}-\vec{y},\vec{r}}
\;\bar{d}_a(\vec{x}t)\gamma_{5}u_{a}(\vec{x}t)\notag}&&\\
&&[\bar{d}_b(\vec{y}t)\gamma_{5}\gamma_{k}d_{b}(\vec{y}t)
+\bar{u}_b(\vec{y}t)\gamma_{5}\gamma_{k}\,u_{b}(\vec{y}t)]
\end{eqnarray}
where $k=1,2,3$. The relative distance $\vec{r}$ may in principle be used to construct
operators that transform according to an irrep of the hypercubic group. However, the
simplest choice $\vec{r}=\vec{0}$ already leads to a viable operator.
Summing over all spatial directions we thus adopt
\begin{equation}\label{pia11}
O_{\pi a_1}(t)=\sum_{k=1}^{3}O_{\pi^{+}a^{0}_{1};k,\vec{0}\,}(t)
\end{equation}
for this simulation.

The operators (\ref{hyopert}) and (\ref{pia11}) are the basis for calculating 
correlation functions
\begin{equation}\label{corXY}
C_{XY}(t,t_0)=
\langle O_{X}(t) O_{Y}^{\dagger}(t_0)\rangle -\langle O_{X}(t)\rangle\langle O_{Y}^{\dagger}(t_0)\rangle\,.
\end{equation}
Here $X$ and $Y$ stand for $h$ or $\pi a_1$, and thus establish a \mbox{$2\times2$} correlation matrix.
The separable terms in (\ref{corXY}) are zero because of the quark flavor assignment in $\h$ and
$\pi a_1$.
The remaining (non-separable) terms in (\ref{corXY}) contain contractions between quark fields at
equal times when worked out with Wick's theorem.
For example, showing flavor structure only,  
\begin{equation}
C_{h,\pi a_1}(t,t_0) \sim\langle(\bar{d}u)_t (\bar{d}d\bar{u}d+\bar{u}u\bar{u}d)_{t_0}\rangle
\end{equation}
with time arguments $t$ and $t_0$ as indicated. Equal-time contractions
:$d\bar{d}$: and :$u\bar{u}$: occur only at the source time slice $t_0$.
The corresponding propagator elements $Q^{-1}(\vec{x}t_{0},\vec{y}t_0)$
are calculated by default. This is different for 
\begin{equation}\label{Cpia1}
C_{\pi a_1,\pi a_1}(t,t_0) \sim\langle(\bar{d}u\bar{d}d+\bar{d}u\bar{u}u)_t
(\bar{d}d\bar{u}d+\bar{u}u\bar{u}d)_{t_0}\rangle
\end{equation}
where we encounter equal time contractions :$d\bar{d}$: and :$u\bar{u}$: at $t > t_0$.
The computation of $Q^{-1}(\vec{x}t,\vec{y}t)$ is very resource intensive and,
if stochastic estimation is used \cite{Michael:1998sg}, then it contributes additional noise.

We shall now argue that this problem can be circumvented:
At $m_d=m_u$ the contractions
:$d\bar{d}$: and :$u\bar{u}$: give rise to the same propagator elements $Q^{-1}(\vec{x}t,\vec{y}t)$.
Thus, replacing the $(\ldots)_t$ term in (\ref{Cpia1}) by $(2\bar{d}u\bar{d}d)_{t}$ and 
reinstating the $\gamma$-matrices from (\ref{piaone})
we observe that $\bar{d}d\sim\bar{d}\gamma_{5}\gamma_{k}d$ couples to $a_1$ and $f_1$ mesons.
Their masses are close however, 1230MeV and 1282MeV respectively \cite{PDBook:2004}.
Invoking a similar argument, altering the quark flavor $d\rightarrow s$ in the above operator
entails
$\bar{d}\gamma_5\gamma_kd\longrightarrow\bar{d}\gamma_5\gamma_ks\sim K_{1}$
and, again, should not significantly alter the mass spectra because
the $K_1$ meson mass of 1270MeV \cite{PDBook:2004} again is close to that of the $a_1$ meson.
In terms of (\ref{Cpia1}) the effect is
\begin{equation}
C_{\pi a_1,\pi a_1}(t,t_0)\rightarrow
\langle(2\bar{d}u\bar{d}s)_{t}(2\bar{s}d\bar{u}d)_{t_0}\rangle\,,\notag
\end{equation}
which now has no equal time contractions, but otherwise is not different from
(\ref{Cpia1}) when worked out in terms of quark propagators.
Hence, dropping equal time contractions in (\ref{Cpia1})
should have little effect on the mass spectrum of the $\pi a_{1}$ system,
and ultimately, on the resulting decay width, because L{\"u}scher's method
for computing scattering phase shifts exclusively relies on the mass spectra, in a finite box.

We emphasize that the correlator element (\ref{Cpia1}) is worked out using the
quark flavor structure exactly as it emerges from (\ref{piaone}) except that equal
time contractions are neglected. All other matrix elements are not effected.

Finally, we do not explicitly compute $C_{\pi a_1,h}(t,t_0)$ but rather
infer it from the hermiticity of the correlation matrix.

\section{Analysis}

For every operator used in this simulation up to three iterations of quark field
smearing \cite{Alexandrou:1994ti} and gauge field fuzzing \cite{Alb87a}
were employed, using 2.5 as a strength parameter in both cases.
The $2\times2$ correlation matrix (\ref{corXY}) thus
expands to size $6\times6$,
\begin{equation}\label{calC}
{\mathcal C}(t,t_0)=\begin{pmatrix}
C_{\h\{ \},\h\{ \}}(t,t_0) & C_{\h\{ \},\pi a_1\{ \}}(t,t_0) \\
C_{\pi a_1\{ \},\h\{ \}}(t,t_0) & C_{\pi a_1\{ \},\pi a_1\{ \}}(t,t_0)\end{pmatrix}\,,
\end{equation}
where the entries are $3\times 3$ matrices with elements
$C_{X\{k\},Y\{\ell\}}(t,t_0)$ built from operators $O_{X\{k\}}(t,t_0)$, etc, with $k=1,2,3$ levels
of fuzzing and smearing.
The latter is done identically at both source and sink, and thus the matrix ${\mathcal C}(t,t_0)$
is hermitian by construction.

A standard analysis method
is based on solving the generalized eigenvalue problem \cite{Luscher:1990ck}
\begin{equation}\label{eigen}
{\mathcal C}(t,t_0)\Psi(t)={\mathcal C}(t_1,t_0)\Psi(t)\Lambda(t)
\end{equation}
where $t_1$ is fixed, $\Psi(t)$ is an $N\times N$ matrix, its columns being the generalized 
eigenvectors, and $\Lambda(t)$ is real diagonal.
We further require that ${\mathcal C}(t_1,t_0)$ be positive 
definite. To ensure the latter $t_1$ should be an ``early" time slice, here we use
$t_{1}-t_{0}=4$. The generalized eigenvalue problem (\ref{eigen}) can then be cast into 
an ordinary one by first diagonalizing 
\begin{equation}\label{ceigen} 
{\mathcal C}(t_1,t_0) = V(t_1,t_0) D(t_1,t_0) V^{\dag}(t_1,t_0)\,,
\end{equation}
where $V(t_1,t_0)$ is unitary and $D(t_1,t_0)$ is real diagonal and positive definite.
Inserting (\ref{ceigen}) into (\ref{eigen}) we are lead to define
\begin{equation}\label{Chatei}
\widehat{C}(t)=\frac{1}{\sqrt{D(t_1,t_0)}}
V^{\dag}(t_1,t_0) {\mathcal C}(t,t_0) V(t_1,t_0) \frac{1}{\sqrt{D(t_1,t_0)}}\,.
\end{equation}
It's eigenvalues are the same as those of the generalized problem (\ref{eigen}), 
$\Lambda(t)=\rm{diag}(\lambda_{1}(t)\ldots\lambda_{N}(t))$.
Constructing $\widehat{C}(t)$ merely amounts to
a linear transformation among the set of operators $O_{X\{k\}}(t,t_0)$ that define
the correlation matrix.
Because of $\widehat{C}(t_1)=\openone$, the transformed set of operators
create quantum states that are orthogonal and normalized at $t=t_1$. Provided that these 
match the ``true" states of the theory, the eigenvectors of $\widehat{C}(t)$
will stay orthogonal as $t\ge t_1$ increases to the extent
allowed by the errors of the simulation. Consequently $\widehat{C}(t)$ will
be diagonal dominated for $t\ge t_1$.
 
\begin{figure}
\includegraphics[height=82mm,angle=90]{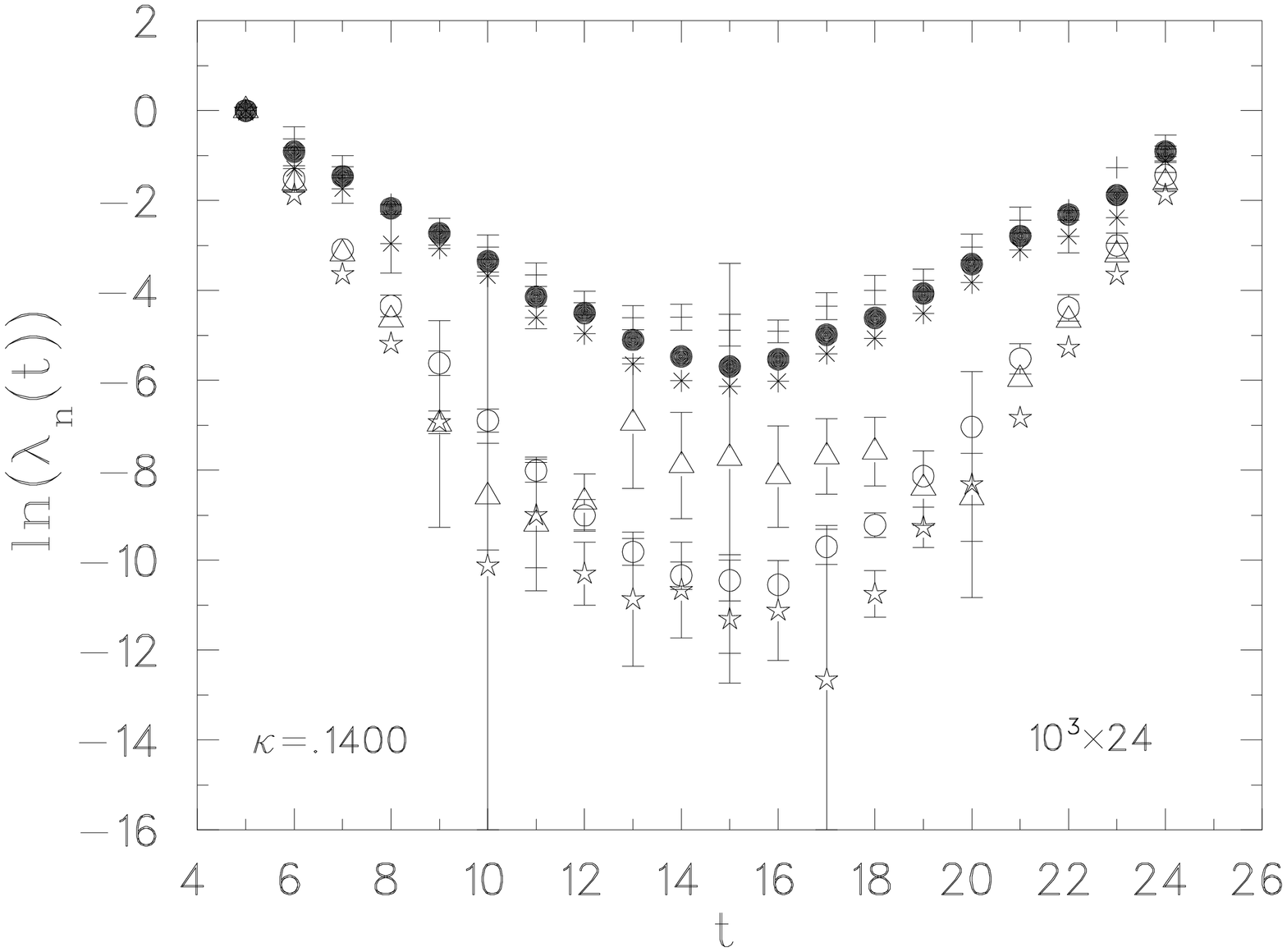}
\includegraphics[height=82mm,angle=90]{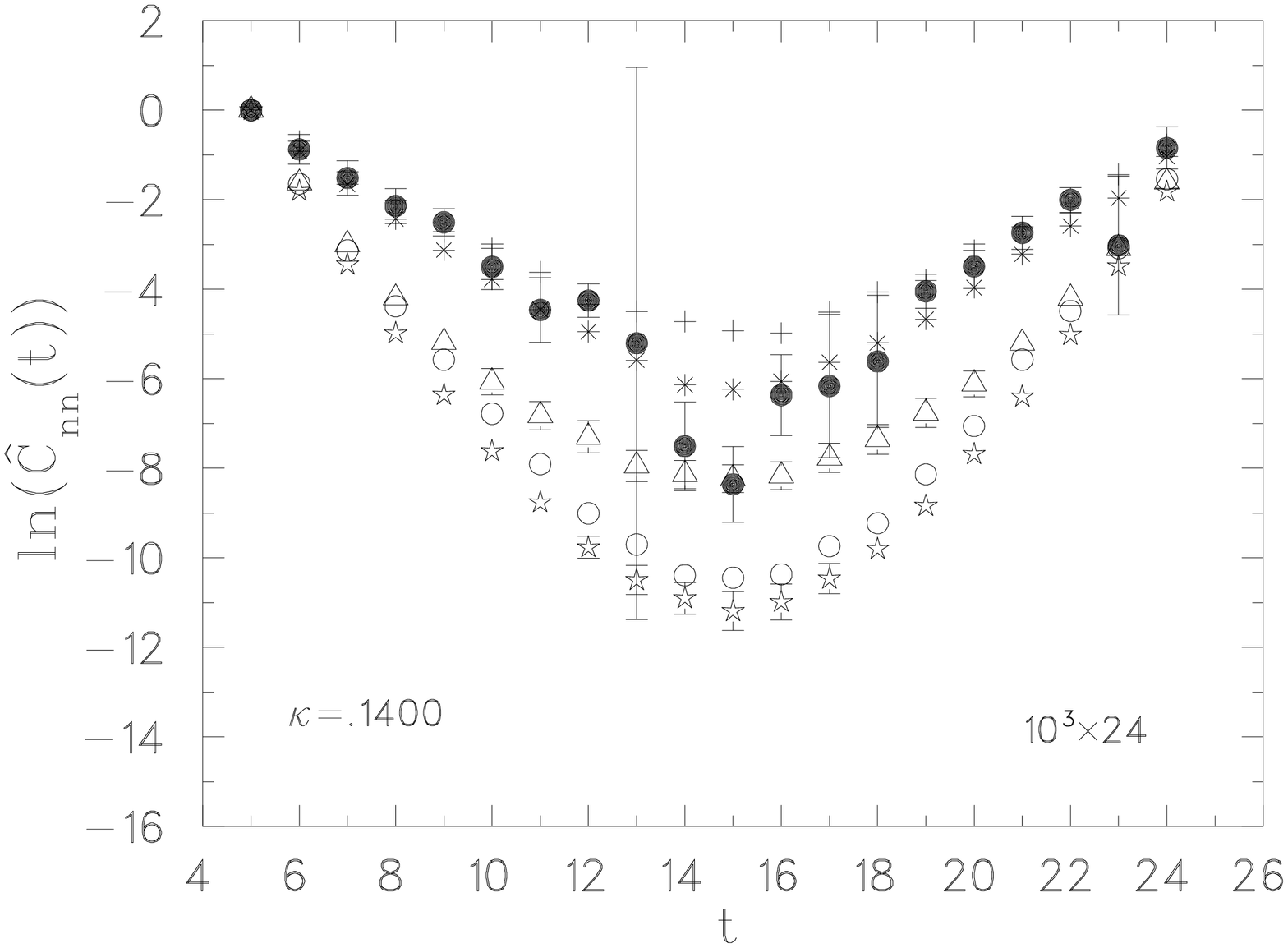}
\caption{\label{comparediag}Comparison of eigenvalues obtained from
diagonalizing $\widehat{C}(t)$ on every time slice
(upper panel) versus projecting to time slice $t=5$ (lower panel), with $t_0=1$.
Results are for the $10^3\times24$ lattice at the lightest pion mass. Fluctuations
are much reduced using the projection technique.}
\end{figure}

From an analysis point of view there are now two options: One may diagonalize
$\widehat{C}(t)$ on each time slice $t$ separately and thus obtain the eigenvalues 
$\lambda_{n}(t),\,n=1 \ldots N$, as proposed in \cite{Luscher:1990ck}.
Alternatively, the diagonal elements of $\widehat{C}(t)$ may 
be taken as an approximation to its eigenvalues,
\begin{equation}\label{project}
\lambda_{n}(t)\approx \widehat{ C}_{nn}(t)\ ,\ n=1 \ldots N\,,
\end{equation}
which involves projecting $\widehat{C}(t)$ into the eigenspaces at fixed
time slice $t_1$, see (\ref{Chatei}).
The latter approach has the advantage that statistical fluctuations
are reduced, see Fig.~\ref{comparediag} for a comparison.
This is plausible because fluctuations of the eigenvector components are effectively frozen.
It also has the advantage to tag the eigenvalues to a specific eigenvector,
which is important for tracking the quark mass dependence of the spectral levels.

\begin{figure}
\includegraphics[height=82mm,angle=90]{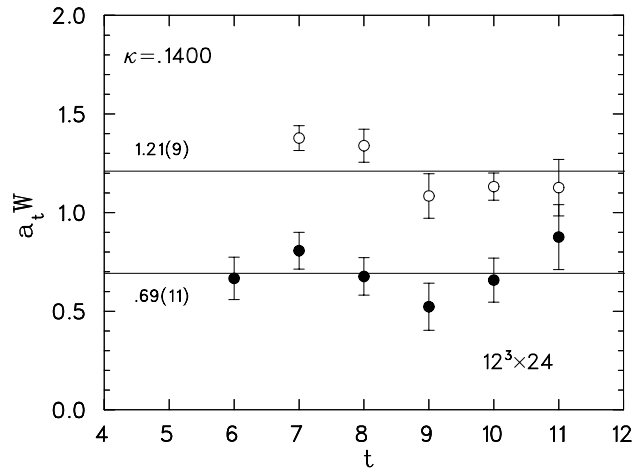}
\caption{\label{efftwo}Example of effective mass functions for two (projected) eigenvalues
of the correlation matrix $\widehat{C}(t)$ on the
$12^3 \times 24$ lattice at the lightest pion mass. The higher mass comes from $\lambda_5(t)$ and
the lower mass from $\lambda_2(t)$.}
\end{figure}

Effective mass function plateaus typically develop in the time interval
$5\lesssim t-t_0 \lesssim 10$, or so. In this region
the projection technique yields more stable results, particularly for the
excited states. An example for the $12^3 \times 24$ lattice at
$\kappa=0.140$ is shown in Fig.~\ref{efftwo}.

Thus we will continue our analysis with the projected 
correlators and, for simplicity, refer to $\widehat{C}_{nn}(t)$ as eigenvalues $\lambda_{n}(t)$.
Those then give rise to the spectral energies $W_n,n=1\dots 6$, listed in
Tab.~\ref{tab:enerspec2}.
\begin{table}[h]
\caption{\label{tab:enerspec2}Energy spectra $W_n, n=1\ldots 6$ from the $12^{3}\times24$
lattice (upper table) and the $10^{3}\times24$ lattice (lower table) at four pion masses.}
\begin{ruledtabular}
\begin{tabular} {lllll}
$\kappa$ & 0.140 & 0.136 & 0.132 & 0.128 \\
\colrule
$a_{t}W_{1}$ & 0.55(14) & 0.69(2)  & 0.74(4)  & 0.82(10) \\
$a_{t}W_{2}$ & 0.69(11) & 0.79(3)  & 0.89(4)  & 0.93(9)  \\
$a_{t}W_{3}$ & 0.81(10) & 0.81(4)  & 0.91(4)  & 1.03(7)  \\
$a_{t}W_{4}$ & 0.88(13) & 1.03(11) & 1.18(10) & 1.26(5)  \\
$a_{t}W_{5}$ & 1.21(9)  & 1.13(3)  & 1.25(3)  & 1.31(4)  \\
$a_{t}W_{6}$ & 0.96(5)  & 1.42(4)  & 1.62(3)  & 1.85(4)  \\
\colrule
$a_{t}W_{1}$ & 0.57(7)  & 0.69(11)  & 0.75(11) & 0.78(6)  \\
$a_{t}W_{2}$ & 0.59(11) & 0.71(5)   & 0.78(4)  & 0.81(5)  \\
$a_{t}W_{3}$ & 0.61(6)  & 0.74(5)   & 0.80(4)  & 0.91(5)  \\
$a_{t}W_{4}$ & 0.91(11) & 1.12(8)   & 1.14(13) & 1.24(9)  \\
$a_{t}W_{5}$ & 1.20(9)  & 1.42(11) & 1.64(9)  & 1.78(9)  \\
$a_{t}W_{6}$ & 1.31(6)  & 1.43(6)   & 1.55(9)  & 1.62(13) \\
\end{tabular}
\end{ruledtabular}
\end{table}

\begin{figure}
\includegraphics[height=82mm,angle=90]{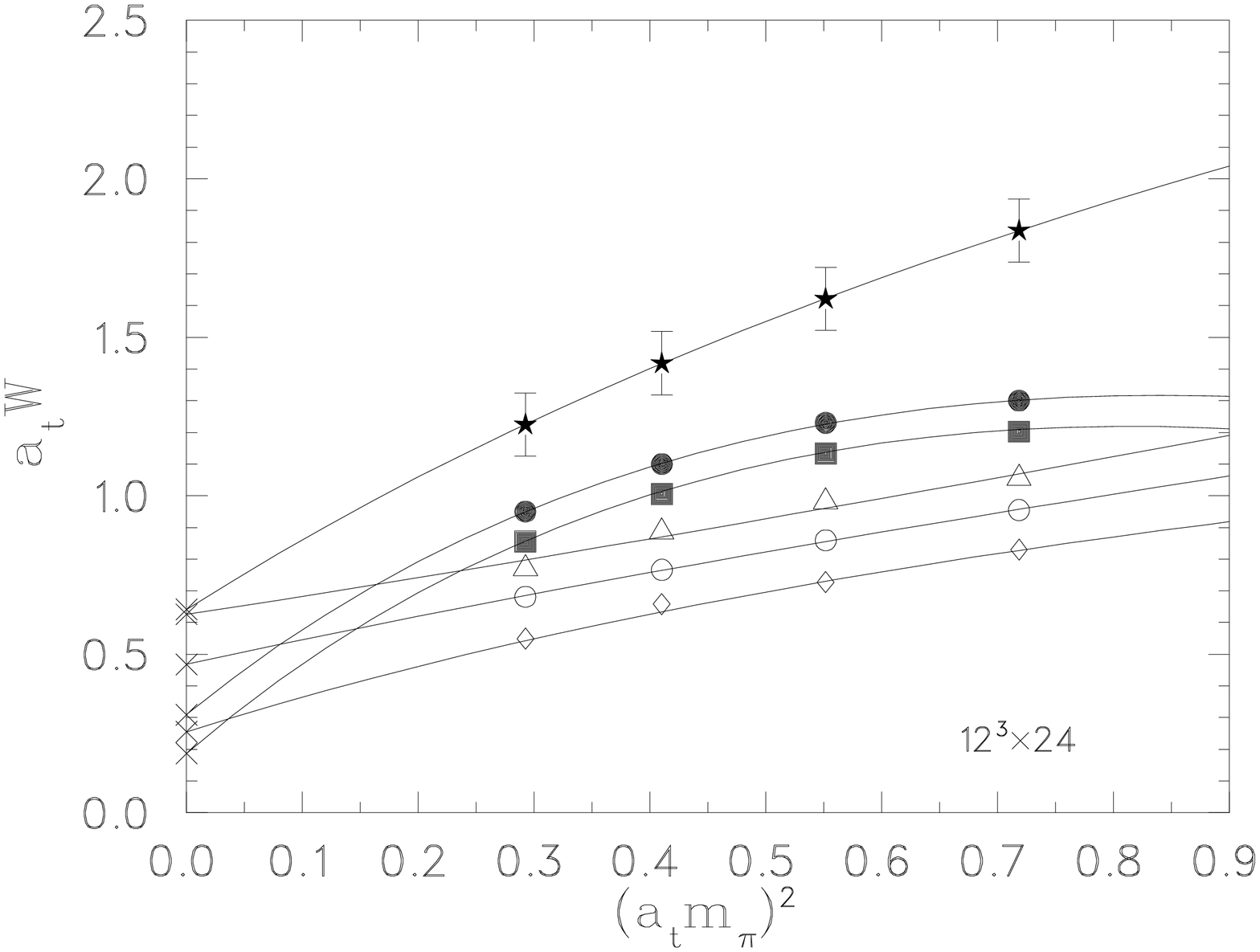}
\includegraphics[height=82mm,angle=90]{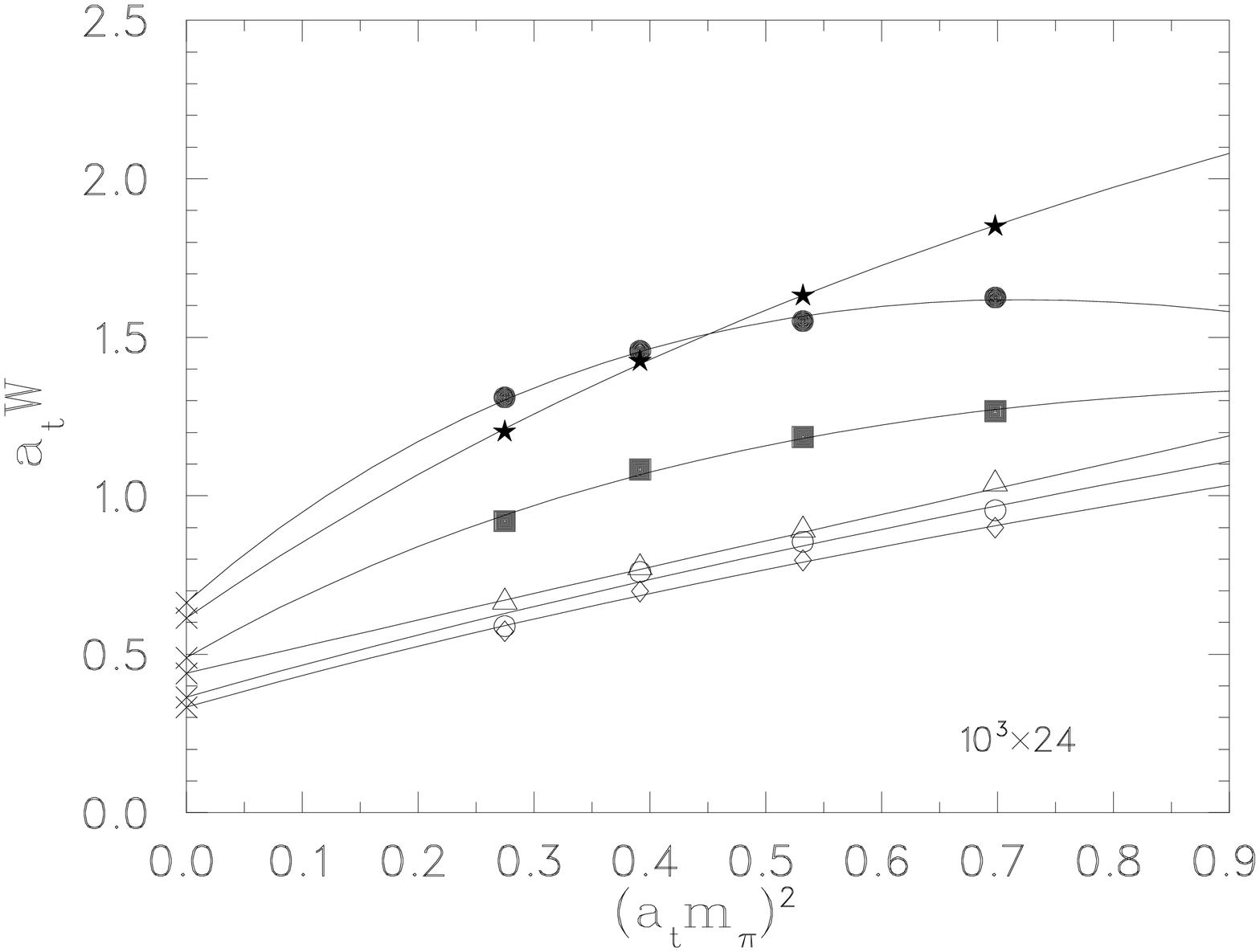}
\caption{\label{spectra}Mass spectra obtained from the $12^3 \times 24$ and $10^3 \times 24$
lattices versus $x=(a_tm_\pi)^2$. Error bars are omitted for clarity except for one level where
the errors shown are typical of all levels.
Fits with the model (\protect\ref{fitln}) are shown as lines.}
\end{figure}

Fits to those spectra with the model (\ref{fitln}) are shown in Fig.~\ref{spectra}.
This figure sheds light on the volume dependence of the spectral levels.
Evidently the ground state mass is relatively stable against changing the lattice volume.
On the other hand the effect on excited states is clearly significant, even to the extent
that level crossing patterns differ for some of the states.
This should not be surprising because excited levels are likely to describe two-meson
states, which are spatially large.
Nevertheless, anticipating results, the volume effect on the scattering phase shift
ultimately turns out to be only moderate. 

Another comment on Fig.~\ref{spectra} is that, although the $6\times 6$ correlation matrix
gives rise to six eigenvalues, the number of physical states
on the lattice
is likely to be a lesser number because, typically, hadronic level spacings are of the order of
a few hundred MeV. Thus we entertain the possibility that the group of the three lower levels
in Fig.~\ref{spectra} describe the same state, the ground state, whereas the upper levels
belong to two-meson states with some degree of interaction energy due to their relative motion.
This point will become more plausible in terms of the corresponding scattering phase shifts.

\section{Scattering phase shifts}

The spectral energies $W_n$ computed on the lattice give rise to a discrete set of relative
$\pi$+$a_1$ momenta $k_n$ by solving the relativistic dispersion relation
\begin{equation}\label{reldisper}
W_{n}=\sqrt{m_{\pi}^2+k^{2}_{n}}+\sqrt{m_{a_1}^2+k^{2}_{n}}\,.
\end{equation}
Note that the resulting momenta $k_n$ relate to spectral masses and thus are continuous
numbers (not subject to lattice discretization).
Only those levels $W_{n}$ will be used that fall within the elastic region,
\begin{equation}\label{welastic}
(m_{\pi}+m_{a_1})< W_{n} < 2 \, (m_{\pi}+m_{a_1})\,.
\end{equation}
Continuum S-wave scattering phase shifts $\delta_0(k)$ are then computed at a discrete
set of momenta $k_n$ using L{\"u}scher's formula \cite{Luscher:1991cf},
\begin{equation}
\tan\delta_0(k_n)=-\frac{\pi^{3/2}q_{n}}{{\cal Z}(1;q_{n}^2)} \quad , \quad q_{n}=
\frac{k_{n}L_{s}}{2\pi}\,.
\end{equation}
Here ${\cal Z}(1;q^2)$ is a generalized $\zeta$-function,
and $L_s=La_s$ is the physical size of the spatial box, using the bare anisotropy $a_s=2a_t$.

If the number of available data points is sufficient, then one may attempt a fit
to a Breit-Wigner function \cite{Tay72},
\begin{eqnarray}
\label{BWmodel}
\tan\delta_0(k)&=&\frac{\Gamma/2}{E_{0}-W(k)}\\
\label{BWmodel1}
\mbox{where} \quad W(k)&=&\sqrt{m_{\pi}^2+k^{2}}+\sqrt{m_{a_1}^2+k^{2}}\,.
\end{eqnarray}
The resonance energy $E_0$ and the decay width $\Gamma$ are fit parameters. 
However, a successful fit can only be expected if the underlying physics indeed
supports an isolated resonance. 
Such a fit actually fails for all spectra computed at the four pion masses,
or rather $x=(a_tm_\pi)^2$, as they appear in Fig.~\ref{spectra}.
This is not be surprising because those data points are far away from a level
crossing between the $\h$ and the $\pi$+$a_1$ masses as evident
from Fig.~\ref{fig:effa1pionhyb}.
It is necessary to extrapolate the spectral masses to $x=0$ near the level crossing.
The model (\ref{fitln}) has been used for this purpose.
We present the extrapolated spectra in Tab.~\ref{tab:eigenmoment1} along with
the corresponding momenta $k_n$ and scattering phase shifts $\delta_0(k_n)$
for those levels which fall into the elastic region (\ref{welastic}).
\begin{table}[h]
\caption{\label{tab:eigenmoment1}Extrapolated energy spectra $W_{n}$
using the model (\protect\ref{fitln}), resulting momenta $k_n$, and S-wave scattering phase
shifts on lattices $12^{3}\times24$ (upper table) and $10^{3}\times24$ (lower table).
Missing entries for $a_{t}k_n$ and $\delta_0(k_n)$ correspond to energy
levels outside of the elastic region (\protect\ref{welastic}).}
\begin{ruledtabular}
\begin{tabular}{clll}
$n$ & $a_{t}W_{n} $ & $a_{t}k_n$ & $\delta_0(k_n)$ \\
\colrule
1 & 0.64(16) & 0.16(1) & 74.6(11.6) \\
2 & 0.63(25) & 0.15(2) & 65.8(14.1) \\
3 & 0.19(5)  & -- & -- \\
4 & 0.47(11) & 0.03(1) & 2.1(1.8)   \\
5 & 0.31(11) & -- & -- \\
6 & 0.26(5)  & -- & -- \\
\colrule
1 & 0.66(21) & 0.20(2) & 77.1(10.5) \\
2 & 0.61(3)  & 0.16(1) & 50.0(6.1)  \\
3 & 0.49(4)  & 0.06(1) & 6.0(3.1)   \\
4 & 0.44(10) & 0.01(1) & 0.2(0.5)   \\
5 & 0.36(10)  & -- & -- \\
6 & 0.33(4)  & -- & -- \\
\end{tabular}
\end{ruledtabular}
\end{table}

The phase shift data are very sparse and do not alone resolve the functional
form of the fit model, such as Breit-Wigner. In fact, attempts of Levenberg-Marquardt
fits using (\ref{BWmodel})
only returned stable results for the resonance energy parameter $a_t E_0$, while the
width parameter $a_t \Gamma$, being an indicator for a derivative,
was left undetermined due to large standard errors.
Nevertheless, it is evident from Tab.~\ref{tab:eigenmoment1} that the phase shift 
data are clustered around two regions of $a_t k_n$,
namely $\approx 0.15$--$0.16$ and $\approx 0.03$ for the $L=12$ lattice,
and $\approx 0.16$--$0.20$ and $\approx 0.01$--$0.06$ for the $L=10$ lattice.
This suggests that no more than two distinct physical states are uncovered by the
simulation. Under this assumption the data may be analyzed as follows:
Denoting the weighted ($\chi^2={\rm min}$) averages for each of the clustered momenta
by $\bar{k}_{1,2}$, the corresponding energies by $\omega_{1,2}= a_t W(\bar{k}_{1,2})$,
and $\tau_{1,2}={\rm tan}\,\delta_0(\bar{k}_{1,2})$, we obtain a set of two equations
from (\ref{BWmodel}), for each lattice, which are solved exactly by
\begin{eqnarray}
\label{gamma}
a_t \Gamma &=& 2(\omega_1-\omega_2)\frac {\tau_1 \tau_2}{\tau_1+\tau_2} \\
\label{E0}
a_t E_0 &=&\frac{\omega_2 \tau_2-\omega_1 \tau_1}{\tau_2-\tau_1}\,.
\end{eqnarray}
The resulting parameter values are listed in Tab.~\ref{tab:results}.
\begin{table}[b]
\caption{\label{tab:results}Results for decay widths $\Gamma$ and resonant energies
$E_0$ for two lattice sizes $L$ in units of $a_t$, and with a physical scale set by
the $\rho$ meson mass.}
\begin{ruledtabular}
\begin{tabular}{ccccc}
$L$ & $a_t\Gamma$ & $a_tE_0$ & $\Gamma[{\rm MeV}]$ & $E_0[{\rm GeV}]$\\
\colrule
12 &0.012(9)& 0.63(3)&35(26)&1.88(8)\\
10 &0.032(14)&0.62(3)&97(43)&1.84(9)\\
\end{tabular}
\end{ruledtabular}
\end{table}
There, the uncertainties for $a_t \Gamma$ and $a_t E_0$ are computed as follows:
The statistical (jackknife) errors for $a_t k_n$, as they appear in
Tab.~\ref{tab:eigenmoment1}, give rise to errors
$\triangle\bar{k}_{1,2}$ for the weighted momentum averages $\bar{k}_{1,2}$.
Repeating the analysis procedure described above a few thousand times with momenta
$k_{1,2}=\bar{k}_{1,2}+\xi\Delta \bar{k}_{1,2}$,
where $\xi$ is a normal distributed random deviate with variance one, then yields the
uncertainties given in Tab.~\ref{tab:results}.
Also the $a_1$ meson mass, which enters the fit model via (\ref{BWmodel1}), was 
subjected to the same randomization. The errors given in Table \ref{tab:results} are the 
standard deviations resulting from the randomization, they are reminiscent of
statistical errors. Table \ref{tab:results} also 
contains the physical values for the decay widths and the resonance energies using the 
$\rho$ meson to set the mass scale.
Setting the scale with the $a_1$ meson mass, results in widths of 39(29)MeV
and 108(48)MeV for the $12^3 \times 24$ and $10^3 \times 24$ lattices respectively.

\begin{figure}[h]
\includegraphics[width=82mm,angle=0]{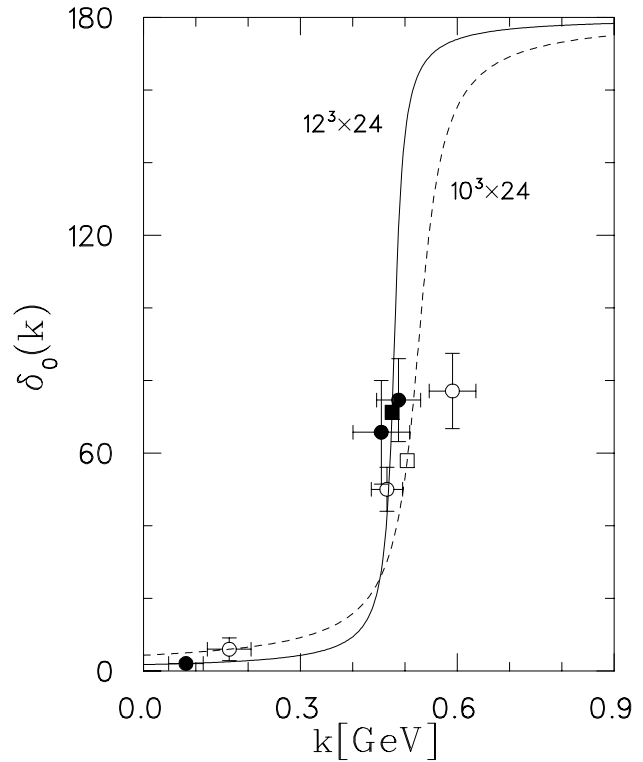}
\caption{\label{scaleindep}Scattering phase shifts $\delta_0(k_{n})$ from
the lattices $12^3 \times 24$ (filled circles) and $10^3 \times 24$ (open circles).
The solid and dashed curves are Breit-Wigner interpolations according to 
(\protect\ref{gamma}) and (\protect\ref{E0}). The filled and open box plot symbols
indicate the respective $\chi^2$-weighted averages over data points.}
\end{figure}

Finally, Fig.~\ref{scaleindep} shows the combined phase shift
data from the two lattices on a common physical scale.
The curves are Breit-Wigner interpolations as explained above.

\section{Errors}

All errors cited in this paper are statistical, and are derived from a standard jackknife
procedure \cite{Efr79}. The hybrid meson operator, involving gauge link paths,
appears to be the major source of those.

The effects of systematic errors on the results of this simulation
are difficult to assess. In principle this can only be done by repeating it with various different
choices of lattice and analysis model parameters. Probably the largest sources
of systematic error stem from curve fitting
and extrapolation techniques. As a check on the extrapolations we have also done fits with
a $x^{3/2}$ term in place of the logarithmic term in (\ref{fitln}).
The results, Tab.~\ref{tab:results}, did not change much, within statistical errors.
If only linear terms are retained the extrapolated masses slightly shift upward, ultimately
resulting in a slight increase of decay widths on the order of $\approx 10$MeV.

Another source of systematic error comes from postulating a Breit-Wigner model.
Given the sparsity 
of data points it is inconclusive that the physical phase shifts will indeed follow a
Breit-Wigner form. In order to resolve this problem the simulation
would have to be repeated at several
values of the gauge coupling $\beta$, thus mapping out some sort of continuous curve $\delta_0(k)$
vs. $k$. The results of this work do rely on the {\it a priori} assumption
that the simulation data follow a Breit-Wigner model.

On the other hand, adopting the less stringent criterion that a resonance is present
if the phase shift data passes through $90^{\circ}$, the simulation results clearly
indicate the presence of such. This, in itself, is a significant outcome of this
project. Although this does not help putting bounds on the systematic error of
$\Gamma$, the results for the resonance energy, $E_0\approx 1.9$ GeV,
are remarkably stable. A decay width, on the other hand, essentially comes from
derivative data and as such is prone to a significantly larger error.

Systematic errors are also caused by finite size effects. At first sight, judging by 
the small difference of the $\rho$ meson masses on the $L=12$ and $L=10$ lattices, see
Table \ref{tab:hopping}, those appear to be small. Finite size effects should be expected
to be much larger for larger-sized hadrons like the $a_1$ for example. This is particularly true
for two-hadron systems studied in this work. For example, the spectra displayed
in Fig.~\ref{spectra}
are significantly different, particularly for excited states on the $L=12$ and $L=10$ lattices,
their size though being quite similar. Again these effects can only be studied by repeating
this simulation with several lattices of different sizes.
    
Obtaining a single scattering phase data point  
requires evaluating up to three effective mass functions - one for the correlator matrix eigenvalue,
one for the $\rho$ meson to set the physical scale, and one for the $\pi$ meson.
The variability in choosing which time slices of the correlation functions 
to use in fitting effective masses produces a variability in the decay width.
Here, one usually wants to maximize the plateau width of the effective mass functions
to optimize the statistical error. Reducing the plateau width to estimate a systematic
error is of limited value.

\section{Conclusion}

Decay widths for the hybrid exotic meson with $J^{PC}=1^{-+}$, calculated using L\"{u}scher's method,
are in the range 35 to 97 MeV with statistical
errors of about 30 MeV using the $\rho$ meson to set the scale. The lower value for the width came 
from using extrapolated energy spectra on a 
$12^3\times 24$ lattice 
and the higher value came from using extrapolated spectra on a $10^3\times 24$ lattice. If the $a_1$
meson sets the scale, then the widths for these two lattices range from 39 to 108 MeV with 
statistical errors of about 40 MeV.

The number of data points available to fit Breit-Wigner functions is very sparse, the reason being
that many energy levels fell outside the elastic region where phase shifts using L\"{u}scher's formulae
cannot be computed. Overcoming this limitation requires use of a larger correlation matrix. This can
be accomplished by adding more smearing levels, and spatially extended operators in the individual
correlators, or possibly by using a coupled channel type analysis in which more than one decay 
channel is represented in the matrix. Several values of the coupling parameter $\beta$ should also
be employed to generate more phase shift data points.  

Using the $\rho$ meson to set the scale, the resonance mass of the hybrid meson in this 
simulation was 1.9(1)~GeV, and in contrast to the decay width, the resonance mass was
well determined 
by the simulation. This unexpected result leads to a final comment that, historically, hadron 
mass calculations within lattice QCD have been done using single-hadron operators, ignoring 
the fact that most hadrons are resonances and thus are unstable \cite{Michael:2005}. We
have taken this decay aspect seriously. Although the numerical values for
the decay widths serve as a guide only, the approach of extracting hadron masses as resonance energies,
 using L\"{u}scher's method, should also be given serious consideration.


\begin{thebibliography}{56}
\expandafter\ifx\csname natexlab\endcsname\relax\def\natexlab#1{#1}\fi
\expandafter\ifx\csname bibnamefont\endcsname\relax
  \def\bibnamefont#1{#1}\fi
\expandafter\ifx\csname bibfnamefont\endcsname\relax
  \def\bibfnamefont#1{#1}\fi
\expandafter\ifx\csname citenamefont\endcsname\relax
  \def\citenamefont#1{#1}\fi
\expandafter\ifx\csname url\endcsname\relax
  \def\url#1{\texttt{#1}}\fi
\expandafter\ifx\csname urlprefix\endcsname\relax\def\urlprefix{URL }\fi
\providecommand{\bibinfo}[2]{#2}
\providecommand{\eprint}[2][]{\url{#2}}

\bibitem[{\citenamefont{Barnes}(2003)}]{Barnes:2003vy}
\bibinfo{author}{\bibfnamefont{T.}~\bibnamefont{Barnes}}
  (\bibinfo{year}{2003}), \eprint{nucl-th/0303032}.

\bibitem[{\citenamefont{Barnes}(2000{\natexlab{a}})}]{Barnes:2000vn}
\bibinfo{author}{\bibfnamefont{T.}~\bibnamefont{Barnes}}
  (\bibinfo{year}{2000}{\natexlab{a}}),
  \eprint[http://arXiv.org/abs]{nucl-th/0009011}.

\bibitem[{\citenamefont{Barnes}(2000{\natexlab{b}})}]{Barnes:2000ns}
\bibinfo{author}{\bibfnamefont{T.}~\bibnamefont{Barnes}},
  \bibinfo{journal}{Acta Phys. Polon.} \textbf{\bibinfo{volume}{B31}},
  \bibinfo{pages}{2545} (\bibinfo{year}{2000}{\natexlab{b}}),
  \eprint{hep-ph/0007296}.

\bibitem[{\citenamefont{Alde et~al.}(1988)}]{Alde:1988bv}
\bibinfo{author}{\bibfnamefont{D.}~\bibnamefont{Alde}} \bibnamefont{et~al.}
  (\bibinfo{collaboration}{IHEP-Brussels-Los Alamos-Annecy (LAPP)}),
  \bibinfo{journal}{Phys. Lett.} \textbf{\bibinfo{volume}{B205}},
  \bibinfo{pages}{397} (\bibinfo{year}{1988}).

\bibitem[{\citenamefont{Thompson et~al.}(1997)}]{Thompson:1997bs}
\bibinfo{author}{\bibfnamefont{D.~R.} \bibnamefont{Thompson}}
  \bibnamefont{et~al.} (\bibinfo{collaboration}{E852}), \bibinfo{journal}{Phys.
  Rev. Lett.} \textbf{\bibinfo{volume}{79}}, \bibinfo{pages}{1630}
  (\bibinfo{year}{1997}), \eprint{hep-ex/9705011}.

\bibitem[{\citenamefont{Abele et~al.}(1998)}]{Abele:1998gn}
\bibinfo{author}{\bibfnamefont{A.}~\bibnamefont{Abele}} \bibnamefont{et~al.}
  (\bibinfo{collaboration}{Crystal Barrel}), \bibinfo{journal}{Phys. Lett.}
  \textbf{\bibinfo{volume}{B423}}, \bibinfo{pages}{175} (\bibinfo{year}{1998}).

\bibitem[{\citenamefont{Adams et~al.}(1998)}]{Adams:1998ff}
\bibinfo{author}{\bibfnamefont{G.~S.} \bibnamefont{Adams}} \bibnamefont{et~al.}
  (\bibinfo{collaboration}{E852}), \bibinfo{journal}{Phys. Rev. Lett.}
  \textbf{\bibinfo{volume}{81}}, \bibinfo{pages}{5760} (\bibinfo{year}{1998}).

\bibitem[{\citenamefont{Chung et~al.}(1999)}]{Chung:1999we}
\bibinfo{author}{\bibfnamefont{S.~U.} \bibnamefont{Chung}} \bibnamefont{et~al.}
  (\bibinfo{collaboration}{E852}), \bibinfo{journal}{Phys. Rev.}
  \textbf{\bibinfo{volume}{D60}}, \bibinfo{pages}{092001}
  (\bibinfo{year}{1999}), \eprint{hep-ex/9902003}.

\bibitem[{\citenamefont{Ivanov et~al.}(2001)}]{Ivanov:2001rv}
\bibinfo{author}{\bibfnamefont{E.~I.} \bibnamefont{Ivanov}}
  \bibnamefont{et~al.} (\bibinfo{collaboration}{E852}), \bibinfo{journal}{Phys.
  Rev. Lett.} \textbf{\bibinfo{volume}{86}}, \bibinfo{pages}{3977}
  (\bibinfo{year}{2001}), \eprint{hep-ex/0101058}.

\bibitem[{\citenamefont{Dzierba et~al.}(2005)}]{Dzierba2005}
\bibinfo{author}{\bibfnamefont{A.~R.} \bibnamefont{Dzierba}}
  \bibnamefont{et~al.} (\bibinfo{year}{2005}), \eprint{hep-ex/0510068}.

\bibitem[{\citenamefont{Barnes et~al.}(1983)\citenamefont{Barnes, Close,
  de~Viron, and Weyers}}]{Barnes:1982tx}
\bibinfo{author}{\bibfnamefont{T.}~\bibnamefont{Barnes}},
  \bibinfo{author}{\bibfnamefont{F.~E.} \bibnamefont{Close}},
  \bibinfo{author}{\bibfnamefont{F.}~\bibnamefont{de~Viron}}, \bibnamefont{and}
  \bibinfo{author}{\bibfnamefont{J.}~\bibnamefont{Weyers}},
  \bibinfo{journal}{Nucl. Phys.} \textbf{\bibinfo{volume}{B224}},
  \bibinfo{pages}{241} (\bibinfo{year}{1983}).

\bibitem[{\citenamefont{Barnes et~al.}(1995)\citenamefont{Barnes, Close, and
  Swanson}}]{Barnes:1995hc}
\bibinfo{author}{\bibfnamefont{T.}~\bibnamefont{Barnes}},
  \bibinfo{author}{\bibfnamefont{F.~E.} \bibnamefont{Close}}, \bibnamefont{and}
  \bibinfo{author}{\bibfnamefont{E.~S.} \bibnamefont{Swanson}},
  \bibinfo{journal}{Phys. Rev.} \textbf{\bibinfo{volume}{D52}},
  \bibinfo{pages}{5242} (\bibinfo{year}{1995}), \eprint{hep-ph/9501405}.

\bibitem[{\citenamefont{Close and Page}(1995)}]{Close:1995hc}
\bibinfo{author}{\bibfnamefont{F.~E.} \bibnamefont{Close}} \bibnamefont{and}
  \bibinfo{author}{\bibfnamefont{P.~R.} \bibnamefont{Page}},
  \bibinfo{journal}{Nucl. Phys.} \textbf{\bibinfo{volume}{B443}},
  \bibinfo{pages}{233} (\bibinfo{year}{1995}),
  \eprint[http://arXiv.org/abs]{hep-ph/9411301}.

\bibitem[{\citenamefont{McNeile and Michael}(2006)}]{McNeile:2006bz}
\bibinfo{author}{\bibfnamefont{C.}~\bibnamefont{McNeile}} \bibnamefont{and}
  \bibinfo{author}{\bibfnamefont{C.}~\bibnamefont{Michael}}
  (\bibinfo{collaboration}{UKQCD}), \bibinfo{journal}{Phys. Rev.}
  \textbf{\bibinfo{volume}{D73}}, \bibinfo{pages}{074506}
  (\bibinfo{year}{2006}), \eprint{hep-lat/0603007}.

\bibitem[{\citenamefont{McNeile et~al.}(2002)\citenamefont{McNeile, Michael,
  and Pennanen}}]{McNeile:2002az}
\bibinfo{author}{\bibfnamefont{C.}~\bibnamefont{McNeile}},
  \bibinfo{author}{\bibfnamefont{C.}~\bibnamefont{Michael}}, \bibnamefont{and}
  \bibinfo{author}{\bibfnamefont{P.}~\bibnamefont{Pennanen}}
  (\bibinfo{collaboration}{UKQCD}), \bibinfo{journal}{Phys. Rev.}
  \textbf{\bibinfo{volume}{D65}}, \bibinfo{pages}{094505}
  (\bibinfo{year}{2002}), \eprint[http://arXiv.org/abs]{hep-lat/0201006}.

\bibitem[{\citenamefont{Michael}(1989)}]{Michael:1989mf}
\bibinfo{author}{\bibfnamefont{C.}~\bibnamefont{Michael}},
  \bibinfo{journal}{Nucl. Phys.} \textbf{\bibinfo{volume}{B327}},
  \bibinfo{pages}{515} (\bibinfo{year}{1989}).

\bibitem[{\citenamefont{Michael}(2005)}]{Michael:2005}
\bibinfo{author}{\bibfnamefont{C.}~\bibnamefont{Michael}}, in
  \emph{\bibinfo{booktitle}{Proceedings of The 23rd International Symposium on
  Lattice Field Theory, 25-30 July 2005, Trinity College, Dublin, Ireland,
  PoS(LAT2005)008}} (\bibinfo{year}{2005}).

\bibitem[{\citenamefont{DeGrand}(1991)}]{Degrand:1991dg}
\bibinfo{author}{\bibfnamefont{T.~A.} \bibnamefont{DeGrand}},
  \bibinfo{journal}{Phys. Rev.} \textbf{\bibinfo{volume}{D43}},
  \bibinfo{pages}{2296} (\bibinfo{year}{1991}).

\bibitem[{\citenamefont{L{\"u}scher}(1991{\natexlab{a}})}]{Luscher:1991cf}
\bibinfo{author}{\bibfnamefont{M.}~\bibnamefont{L{\"u}scher}},
  \bibinfo{journal}{Nucl. Phys.} \textbf{\bibinfo{volume}{B364}},
  \bibinfo{pages}{237} (\bibinfo{year}{1991}{\natexlab{a}}).

\bibitem[{\citenamefont{Lellouch and L{\"u}scher}(2001)}]{Lellouch:2000pv}
\bibinfo{author}{\bibfnamefont{L.}~\bibnamefont{Lellouch}} \bibnamefont{and}
  \bibinfo{author}{\bibfnamefont{M.}~\bibnamefont{L{\"u}scher}},
  \bibinfo{journal}{Commun. Math. Phys.} \textbf{\bibinfo{volume}{219}},
  \bibinfo{pages}{31} (\bibinfo{year}{2001}),
  \eprint[http://arXiv.org/abs]{hep-lat/0003023}.

\bibitem[{\citenamefont{Juge et~al.}(1999)\citenamefont{Juge, Kuti, and
  Morningstar}}]{Juge:1999ie}
\bibinfo{author}{\bibfnamefont{K.~J.} \bibnamefont{Juge}},
  \bibinfo{author}{\bibfnamefont{J.}~\bibnamefont{Kuti}}, \bibnamefont{and}
  \bibinfo{author}{\bibfnamefont{C.~J.} \bibnamefont{Morningstar}},
  \bibinfo{journal}{Phys. Rev. Lett.} \textbf{\bibinfo{volume}{82}},
  \bibinfo{pages}{4400} (\bibinfo{year}{1999}),
  \eprint[http://arXiv.org/abs]{hep-ph/9902336}.

\bibitem[{\citenamefont{Juge et~al.}(1998)\citenamefont{Juge, Kuti, and
  Morningstar}}]{Juge:1998nc}
\bibinfo{author}{\bibfnamefont{K.~J.} \bibnamefont{Juge}},
  \bibinfo{author}{\bibfnamefont{J.}~\bibnamefont{Kuti}}, \bibnamefont{and}
  \bibinfo{author}{\bibfnamefont{C.~J.} \bibnamefont{Morningstar}},
  \bibinfo{journal}{Nucl. Phys. Proc. Suppl.} \textbf{\bibinfo{volume}{63}},
  \bibinfo{pages}{326} (\bibinfo{year}{1998}),
  \eprint[http://arXiv.org/abs]{hep-lat/9709131}.

\bibitem[{\citenamefont{Perantonis and Michael}(1990)}]{Perantonis:1990dy}
\bibinfo{author}{\bibfnamefont{S.}~\bibnamefont{Perantonis}} \bibnamefont{and}
  \bibinfo{author}{\bibfnamefont{C.}~\bibnamefont{Michael}},
  \bibinfo{journal}{Nucl. Phys.} \textbf{\bibinfo{volume}{B347}},
  \bibinfo{pages}{854} (\bibinfo{year}{1990}).

\bibitem[{\citenamefont{Burch et~al.}(2001)\citenamefont{Burch, Orginos, and
  Toussaint}}]{Burch:2001tr}
\bibinfo{author}{\bibfnamefont{T.}~\bibnamefont{Burch}},
  \bibinfo{author}{\bibfnamefont{K.}~\bibnamefont{Orginos}}, \bibnamefont{and}
  \bibinfo{author}{\bibfnamefont{D.}~\bibnamefont{Toussaint}},
  \bibinfo{journal}{Phys. Rev.} \textbf{\bibinfo{volume}{D64}},
  \bibinfo{pages}{074505} (\bibinfo{year}{2001}),
  \eprint[http://arXiv.org/abs]{hep-lat/0103025}.

\bibitem[{\citenamefont{Drummond et~al.}(2000)\citenamefont{Drummond, Goodman,
  Horgan, Shanahan, and Storoni}}]{Drummond:1999db}
\bibinfo{author}{\bibfnamefont{I.~T.} \bibnamefont{Drummond}},
  \bibinfo{author}{\bibfnamefont{N.~A.} \bibnamefont{Goodman}},
  \bibinfo{author}{\bibfnamefont{R.~R.} \bibnamefont{Horgan}},
  \bibinfo{author}{\bibfnamefont{H.~P.} \bibnamefont{Shanahan}},
  \bibnamefont{and} \bibinfo{author}{\bibfnamefont{L.~C.}
  \bibnamefont{Storoni}}, \bibinfo{journal}{Phys. Lett.}
  \textbf{\bibinfo{volume}{B478}}, \bibinfo{pages}{151} (\bibinfo{year}{2000}),
  \eprint[http://arXiv.org/abs]{hep-lat/9912041}.

\bibitem[{\citenamefont{Manke et~al.}(1998)\citenamefont{Manke, Drummond,
  Horgan, and Shanahan}}]{Manke:1998tb}
\bibinfo{author}{\bibfnamefont{T.}~\bibnamefont{Manke}},
  \bibinfo{author}{\bibfnamefont{I.~T.} \bibnamefont{Drummond}},
  \bibinfo{author}{\bibfnamefont{R.~R.} \bibnamefont{Horgan}},
  \bibnamefont{and} \bibinfo{author}{\bibfnamefont{H.~P.}
  \bibnamefont{Shanahan}} (\bibinfo{collaboration}{UKQCD}),
  \bibinfo{journal}{Nucl. Phys. Proc. Suppl.} \textbf{\bibinfo{volume}{63}},
  \bibinfo{pages}{332} (\bibinfo{year}{1998}),
  \eprint[http://arXiv.org/abs]{hep-lat/9709001}.

\bibitem[{\citenamefont{McNeile et~al.}(1999)}]{McNeile:1998cp}
\bibinfo{author}{\bibfnamefont{C.}~\bibnamefont{McNeile}} \bibnamefont{et~al.},
  \bibinfo{journal}{Nucl. Phys. Proc. Suppl.} \textbf{\bibinfo{volume}{73}},
  \bibinfo{pages}{264} (\bibinfo{year}{1999}),
  \eprint[http://arXiv.org/abs]{hep-lat/9809087}.

\bibitem[{\citenamefont{Bernard et~al.}(1997{\natexlab{a}})}]{Bernard:1997ib}
\bibinfo{author}{\bibfnamefont{C.~W.} \bibnamefont{Bernard}}
  \bibnamefont{et~al.} (\bibinfo{collaboration}{MILC}), \bibinfo{journal}{Phys.
  Rev.} \textbf{\bibinfo{volume}{D56}}, \bibinfo{pages}{7039}
  (\bibinfo{year}{1997}{\natexlab{a}}),
  \eprint[http://arXiv.org/abs]{hep-lat/9707008}.

\bibitem[{\citenamefont{Bernard et~al.}(1997{\natexlab{b}})}]{Bernard:1997bg}
\bibinfo{author}{\bibfnamefont{C.~W.} \bibnamefont{Bernard}}
  \bibnamefont{et~al.}, \bibinfo{journal}{Nucl. Phys. Proc. Suppl.}
  \textbf{\bibinfo{volume}{53}}, \bibinfo{pages}{228}
  (\bibinfo{year}{1997}{\natexlab{b}}),
  \eprint[http://arXiv.org/abs]{hep-lat/9607031}.

\bibitem[{\citenamefont{Lacock and Schilling}(1999)}]{Lacock:1998be}
\bibinfo{author}{\bibfnamefont{P.}~\bibnamefont{Lacock}} \bibnamefont{and}
  \bibinfo{author}{\bibfnamefont{K.}~\bibnamefont{Schilling}}
  (\bibinfo{collaboration}{TXL}), \bibinfo{journal}{Nucl. Phys. Proc. Suppl.}
  \textbf{\bibinfo{volume}{73}}, \bibinfo{pages}{261} (\bibinfo{year}{1999}),
  \eprint[http://arXiv.org/abs]{hep-lat/9809022}.

\bibitem[{\citenamefont{Lacock et~al.}(1998)\citenamefont{Lacock, Michael,
  Boyle, and Rowland}}]{Lacock:1998an}
\bibinfo{author}{\bibfnamefont{P.}~\bibnamefont{Lacock}},
  \bibinfo{author}{\bibfnamefont{C.}~\bibnamefont{Michael}},
  \bibinfo{author}{\bibfnamefont{P.}~\bibnamefont{Boyle}}, \bibnamefont{and}
  \bibinfo{author}{\bibfnamefont{P.}~\bibnamefont{Rowland}}
  (\bibinfo{collaboration}{UKQCD}), \bibinfo{journal}{Nucl. Phys. Proc. Suppl.}
  \textbf{\bibinfo{volume}{63}}, \bibinfo{pages}{203} (\bibinfo{year}{1998}),
  \eprint[http://arXiv.org/abs]{hep-lat/9708013}.

\bibitem[{\citenamefont{Lacock et~al.}(1997)\citenamefont{Lacock, Michael,
  Boyle, and Rowland}}]{Lacock:1997ny}
\bibinfo{author}{\bibfnamefont{P.}~\bibnamefont{Lacock}},
  \bibinfo{author}{\bibfnamefont{C.}~\bibnamefont{Michael}},
  \bibinfo{author}{\bibfnamefont{P.}~\bibnamefont{Boyle}}, \bibnamefont{and}
  \bibinfo{author}{\bibfnamefont{P.}~\bibnamefont{Rowland}}
  (\bibinfo{collaboration}{UKQCD}), \bibinfo{journal}{Phys. Lett.}
  \textbf{\bibinfo{volume}{B401}}, \bibinfo{pages}{308} (\bibinfo{year}{1997}),
  \eprint[http://arXiv.org/abs]{hep-lat/9611011}.

\bibitem[{\citenamefont{Lacock et~al.}(1996)\citenamefont{Lacock, Michael,
  Boyle, and Rowland}}]{Lacock:1996vy}
\bibinfo{author}{\bibfnamefont{P.}~\bibnamefont{Lacock}},
  \bibinfo{author}{\bibfnamefont{C.}~\bibnamefont{Michael}},
  \bibinfo{author}{\bibfnamefont{P.}~\bibnamefont{Boyle}}, \bibnamefont{and}
  \bibinfo{author}{\bibfnamefont{P.}~\bibnamefont{Rowland}}
  (\bibinfo{collaboration}{UKQCD}), \bibinfo{journal}{Phys. Rev.}
  \textbf{\bibinfo{volume}{D54}}, \bibinfo{pages}{6997} (\bibinfo{year}{1996}),
  \eprint[http://arXiv.org/abs]{hep-lat/9605025}.

\bibitem[{\citenamefont{Bernard~{\it et al}}(2003)}]{Bernard:2003cw}
\bibinfo{author}{\bibfnamefont{C.}~\bibnamefont{Bernard~{\it et al}}}
  (\bibinfo{collaboration}{MILC}), \bibinfo{journal}{Nucl. Phys. B (Proc.
  Suppl.)} \textbf{\bibinfo{volume}{119}}, \bibinfo{pages}{260}
  (\bibinfo{year}{2003}), \eprint[http://arXiv.org/abs]{hep-lat/0209097}.

\bibitem[{\citenamefont{McNeile and Michael}(2003)}]{McNeile:2002fh}
\bibinfo{author}{\bibfnamefont{C.}~\bibnamefont{McNeile}} \bibnamefont{and}
  \bibinfo{author}{\bibfnamefont{C.}~\bibnamefont{Michael}}
  (\bibinfo{collaboration}{UKQCD}), \bibinfo{journal}{Phys. Lett.}
  \textbf{\bibinfo{volume}{B556}}, \bibinfo{pages}{177} (\bibinfo{year}{2003}),
  \eprint{hep-lat/0212020}.

\bibitem[{\citenamefont{L{\"u}scher}(1991{\natexlab{b}})}]{Luscher:1991ux}
\bibinfo{author}{\bibfnamefont{M.}~\bibnamefont{L{\"u}scher}},
  \bibinfo{journal}{Nucl. Phys.} \textbf{\bibinfo{volume}{B354}},
  \bibinfo{pages}{531} (\bibinfo{year}{1991}{\natexlab{b}}).

\bibitem[{\citenamefont{Cook}(2006)}]{Cook:2006mc}
\bibinfo{author}{\bibfnamefont{M.~S.} \bibnamefont{Cook}},
  \emph{\bibinfo{title}{Exotic Meson Decay Widths using Lattice Quantum
  Chromodynamics}} (\bibinfo{publisher}{Ph.D. Dissertation, Florida
  International University}, \bibinfo{address}{Miami, Florida},
  \bibinfo{year}{2006}).

\bibitem[{\citenamefont{L{\"u}scher and Wolff}(1990)}]{Luscher:1990ck}
\bibinfo{author}{\bibfnamefont{M.}~\bibnamefont{L{\"u}scher}} \bibnamefont{and}
  \bibinfo{author}{\bibfnamefont{U.}~\bibnamefont{Wolff}},
  \bibinfo{journal}{Nucl. Phys.} \textbf{\bibinfo{volume}{B339}},
  \bibinfo{pages}{222} (\bibinfo{year}{1990}).

\bibitem[{\citenamefont{G{\"o}ckeler et~al.}(1994)\citenamefont{G{\"o}ckeler,
  Kastrup, Westphalen, and Zimmermann}}]{Gockeler:1994rx}
\bibinfo{author}{\bibfnamefont{M.}~\bibnamefont{G{\"o}ckeler}},
  \bibinfo{author}{\bibfnamefont{H.~A.} \bibnamefont{Kastrup}},
  \bibinfo{author}{\bibfnamefont{J.}~\bibnamefont{Westphalen}},
  \bibnamefont{and}
  \bibinfo{author}{\bibfnamefont{F.}~\bibnamefont{Zimmermann}},
  \bibinfo{journal}{Nucl. Phys.} \textbf{\bibinfo{volume}{B425}},
  \bibinfo{pages}{413} (\bibinfo{year}{1994}),
  \eprint[http://arXiv.org/abs]{hep-lat/9402011}.

\bibitem[{\citenamefont{Zimmermann et~al.}(1993)\citenamefont{Zimmermann,
  Westphalen, G{\"o}ckeler, and Kastrup}}]{Zimmermann:1993kx}
\bibinfo{author}{\bibfnamefont{F.}~\bibnamefont{Zimmermann}},
  \bibinfo{author}{\bibfnamefont{J.}~\bibnamefont{Westphalen}},
  \bibinfo{author}{\bibfnamefont{M.}~\bibnamefont{G{\"o}ckeler}},
  \bibnamefont{and} \bibinfo{author}{\bibfnamefont{H.~A.}
  \bibnamefont{Kastrup}}, \bibinfo{journal}{Nucl. Phys. Proc. Suppl.}
  \textbf{\bibinfo{volume}{30}}, \bibinfo{pages}{879} (\bibinfo{year}{1993}),
  \eprint[http://arXiv.org/abs]{hep-lat/9211029}.

\bibitem[{\citenamefont{Fiebig et~al.}(1994)\citenamefont{Fiebig, Dominguez,
  and Woloshyn}}]{Fiebig:1994qi}
\bibinfo{author}{\bibfnamefont{H.~R.} \bibnamefont{Fiebig}},
  \bibinfo{author}{\bibfnamefont{A.}~\bibnamefont{Dominguez}},
  \bibnamefont{and} \bibinfo{author}{\bibfnamefont{R.~M.}
  \bibnamefont{Woloshyn}}, \bibinfo{journal}{Nucl. Phys.}
  \textbf{\bibinfo{volume}{B418}}, \bibinfo{pages}{649} (\bibinfo{year}{1994}).

\bibitem[{Gat({\natexlab{a}})}]{Gat93b}
\bibinfo{note}{C.R. Gattringer and C.B. Lang, Nucl. Phys. B 391 (1993) 463}.

\bibitem[{Gat({\natexlab{b}})}]{Gat93a}
\bibinfo{note}{C.R. Gattringer , I. Hip and C.B. Lang, Nucl.Phys. B (Proc.
  Suppl.) 30 (1993) 875}.

\bibitem[{\citenamefont{Loft and DeGrand}(1989)}]{Loft:1988sy}
\bibinfo{author}{\bibfnamefont{R.~D.} \bibnamefont{Loft}} \bibnamefont{and}
  \bibinfo{author}{\bibfnamefont{T.~A.} \bibnamefont{DeGrand}},
  \bibinfo{journal}{Phys. Rev.} \textbf{\bibinfo{volume}{D39}},
  \bibinfo{pages}{2692} (\bibinfo{year}{1989}).

\bibitem[{\citenamefont{Eidelman et~al.}(2004)}]{PDBook:2004}
\bibinfo{author}{\bibfnamefont{S.}~\bibnamefont{Eidelman}}
  \bibnamefont{et~al.}, \bibinfo{journal}{Physics Letters B}
  \textbf{\bibinfo{volume}{592}}, \bibinfo{pages}{1+} (\bibinfo{year}{2004}),
  \urlprefix\url{http://pdg.lbl.gov}.

\bibitem[{\citenamefont{Alexandrou et~al.}(1994)\citenamefont{Alexandrou,
  G{\"u}sken, Jegerlehner, Schilling, and Sommer}}]{Alexandrou:1994ti}
\bibinfo{author}{\bibfnamefont{C.}~\bibnamefont{Alexandrou}},
  \bibinfo{author}{\bibfnamefont{S.}~\bibnamefont{G{\"u}sken}},
  \bibinfo{author}{\bibfnamefont{F.}~\bibnamefont{Jegerlehner}},
  \bibinfo{author}{\bibfnamefont{K.}~\bibnamefont{Schilling}},
  \bibnamefont{and} \bibinfo{author}{\bibfnamefont{R.}~\bibnamefont{Sommer}},
  \bibinfo{journal}{Nucl. Phys.} \textbf{\bibinfo{volume}{B414}},
  \bibinfo{pages}{815} (\bibinfo{year}{1994}),
  \eprint[http://arXiv.org/abs]{hep-lat/9211042}.

\bibitem[{\citenamefont{Albanese et~al.}(1987)}]{Alb87a}
\bibinfo{author}{\bibfnamefont{C.}~\bibnamefont{Albanese}}
  \bibnamefont{et~al.}, \bibinfo{journal}{Phys.\ Lett.}
  \textbf{\bibinfo{volume}{B192}}, \bibinfo{pages}{163} (\bibinfo{year}{1987}).

\bibitem[{\citenamefont{Bernard~{\it et al}}(1997)}]{Bernard:1997ex}
\bibinfo{author}{\bibfnamefont{C.}~\bibnamefont{Bernard~{\it et al}}},
  \bibinfo{journal}{Phys. Rev.} \textbf{\bibinfo{volume}{D56}},
  \bibinfo{pages}{7040} (\bibinfo{year}{1997}).

\bibitem[{\citenamefont{Frommer et~al.}(1995)\citenamefont{Frommer, N{\"o}ckel,
  G{\"u}sken, Lippert, and Schilling}}]{Frommer:1995ik}
\bibinfo{author}{\bibfnamefont{A.}~\bibnamefont{Frommer}},
  \bibinfo{author}{\bibfnamefont{B.}~\bibnamefont{N{\"o}ckel}},
  \bibinfo{author}{\bibfnamefont{S.}~\bibnamefont{G{\"u}sken}},
  \bibinfo{author}{\bibfnamefont{T.}~\bibnamefont{Lippert}}, \bibnamefont{and}
  \bibinfo{author}{\bibfnamefont{K.}~\bibnamefont{Schilling}},
  \bibinfo{journal}{Int. J. Mod. Phys.} \textbf{\bibinfo{volume}{C6}},
  \bibinfo{pages}{627} (\bibinfo{year}{1995}),
  \eprint[http://arXiv.org/abs]{hep-lat/9504020}.

\bibitem[{Ber(1995)}]{BernsteinHolstein:1995}
in \emph{\bibinfo{booktitle}{Chiral Dynamics: Theory and Experiment}}, edited
  by \bibinfo{editor}{\bibfnamefont{A.~M.} \bibnamefont{Bernstein}}
  \bibnamefont{and} \bibinfo{editor}{\bibfnamefont{B.~R.}
  \bibnamefont{Holstein}} (\bibinfo{publisher}{Springer-Verlag},
  \bibinfo{address}{Berlin, Heidelberg, New York}, \bibinfo{year}{1995}), vol.
  \bibinfo{volume}{452} of \emph{\bibinfo{series}{Lecture Notes in Physics}}.

\bibitem[{\citenamefont{Smit}(2002)}]{Smit:2002sm}
\bibinfo{author}{\bibfnamefont{J.}~\bibnamefont{Smit}},
  \emph{\bibinfo{title}{Introduction to Quantum Fields on a Lattice}}
  (\bibinfo{publisher}{Cambridge University Press},
  \bibinfo{address}{Cambridge,UK}, \bibinfo{year}{2002}).

\bibitem[{\citenamefont{Wright et~al.}(2002)}]{Wright:2002wr}
\bibinfo{author}{\bibfnamefont{S.~V.} \bibnamefont{Wright}}
  \bibnamefont{et~al.}, \bibinfo{journal}{Nucl. Phys. Proc. Suppl}
  \textbf{\bibinfo{volume}{109A}}, \bibinfo{pages}{50} (\bibinfo{year}{2002}),
  \eprint{hep-lat/0111053}.

\bibitem[{\citenamefont{Hedditch et~al.}(2005)}]{Hedditch:2005zf}
\bibinfo{author}{\bibfnamefont{J.~N.} \bibnamefont{Hedditch}}
  \bibnamefont{et~al.}, \bibinfo{journal}{Phys. Rev.}
  \textbf{\bibinfo{volume}{D72}}, \bibinfo{pages}{114507}
  (\bibinfo{year}{2005}), \eprint{hep-lat/0509106}.

\bibitem[{\citenamefont{Michael and Peisa}(1998)}]{Michael:1998sg}
\bibinfo{author}{\bibfnamefont{C.}~\bibnamefont{Michael}} \bibnamefont{and}
  \bibinfo{author}{\bibfnamefont{J.}~\bibnamefont{Peisa}}
  (\bibinfo{collaboration}{UKQCD}), \bibinfo{journal}{Phys. Rev.}
  \textbf{\bibinfo{volume}{D58}}, \bibinfo{pages}{034506}
  (\bibinfo{year}{1998}), \eprint[http://arXiv.org/abs]{hep-lat/9802015}.

\bibitem[{\citenamefont{Taylor}(1972)}]{Tay72}
\bibinfo{author}{\bibfnamefont{J.~R.} \bibnamefont{Taylor}},
  \emph{\bibinfo{title}{Scattering Theory}} (\bibinfo{publisher}{Wiley},
  \bibinfo{address}{New York}, \bibinfo{year}{1972}).

\bibitem[{Efr()}]{Efr79}
\bibinfo{note}{B. Efron, SIAM Review 21 (1979) 460}.

\end{thebibliography}
\end{document}